\documentclass[conference]{IEEEtran}
\IEEEoverridecommandlockouts

\usepackage{amsfonts}
\usepackage{amssymb}
\usepackage{cite}
\usepackage{amsmath}
\usepackage{stfloats}
\usepackage{booktabs}
\usepackage{array}
\usepackage{tabularx}
\usepackage{slashbox}
\usepackage[dvips]{graphicx}
\usepackage{subfigure}
\usepackage{multirow}
\usepackage{bm}
\usepackage{color}
\usepackage{float}

\begin{document}
%
\title{Composite Learning Control With Application\\to Inverted Pendulums}



%
\author{\IEEEauthorblockN{Yongping Pan\IEEEauthorrefmark{1},
Lin Pan\IEEEauthorrefmark{2}\IEEEauthorrefmark{3}, and
Haoyong Yu\IEEEauthorrefmark{1}}
\IEEEauthorblockA{\IEEEauthorrefmark{1}School of Biomedical Engineering, National University of Singapore, Singapore 117575, Singapore
\\\tt\small Email: biepany@nus.edu.sg; bieyhy@nus.edu.sg}
\IEEEauthorblockA{\IEEEauthorrefmark{2}Interdisciplinary Centre for Security, Reliability and Trust, University of Luxembourg, Luxembourg\\
\tt\small Email: lin.pan@uni.lu}
\IEEEauthorblockA{\IEEEauthorrefmark{3}School of Electric and Electronic Engineering, Wuhan Polytechnic University, Wuhan 430023, China}
}


\maketitle

\begin{abstract}
Composite adaptive control (CAC) that integrates direct and indirect adaptive control techniques can achieve smaller tracking errors and faster parameter convergence compared with direct and indirect adaptive control techniques. However, the condition of persistent excitation (PE) still has to be satisfied to guarantee parameter convergence in CAC. This paper proposes a novel model reference composite learning control (MRCLC) strategy for a class of affine nonlinear systems with parametric uncertainties to guarantee parameter convergence without the PE condition. In the composite learning, an integral during a moving-time window is utilized to construct a prediction error, a linear filter is applied to alleviate the derivation of plant states, and both the tracking error and the prediction error are applied to update parametric estimates. It is proven that the closed-loop system achieves global exponential-like stability under interval excitation rather than PE of regression functions. The effectiveness of the proposed MRCLC has been verified by the application to an inverted pendulum control problem.

\end{abstract}


%
\IEEEpeerreviewmaketitle

\section{Introduction}

Adaptive control is one of the major control techniques of handling uncertainties in nonlinear systems and has still attracted great concern in recent years \cite{Astrom2008, Astolfi2008, Anderson2008, Stefanovic2011, Narendra2011, Casavola2012, Khan2012, Martin2012, Pathak2012, Lavretsky2013, Serrani2013, Sun2013, Barkana2014b, Chan2014, Swarnkar2014, Tao2014}. In particular, \emph{model reference adaptive control} (MRAC) is a popular adaptive control architecture which aims to make an uncertain dynamical system behave like a chosen reference model. The way of parameter estimation in adaptive control gives rise to two different schemes, namely indirect and direct schemes \cite{Ioannou1996}. \emph{Composite adaptive control} (CAC) is an integrated direct and indirect adaptive control technique which aims to achieve better tracking and parameter estimation through faster and smoother parameter adaptation \cite{Slotine1989}. In the CAC, prediction errors are generated by identification models, and both tracking errors and prediction errors are applied to update parametric estimates. The superiority of CAC on performance improvement has been demonstrated in many control designs, where some latest results can be referred to \cite{Naso2010, Patre2010, Hu2010, Mohanty2011, Pan2012, Pan2013b, Wei2013, Dydek2013}. Nevertheless, as in the classical adaptive control, CAC only achieves asymptotic convergence of tracking errors and does not guarantee parameter convergence unless plant states satisfy the condition of persistent excitation (PE) \cite{Ioannou1996}. It is well known that the PE condition is very strict and often infeasible to monitor online in practical control systems \cite{Chowdhary2010a}.

Learning is one of the fundamental features of autonomous intelligent behavior, and it is closely related to parameter convergence in adaptive control \cite{Antsaklis1995}. The benefits brought by parameter convergence include accurate closed-loop identification, exponential stability and robustness against measurement noise \cite{Chowdhary2014}. An emerging \emph{concurrent learning} technique provides a feasible and promising way for achieving parameter convergence without the PE condition in MRAC \cite{Chowdhary2010a, Chowdhary2011, Chowdhary2014}. The difference between the concurrent learning and the composite adaptation lies in how to construct prediction errors. In the concurrent learning, a dynamic data stack constituted by online recorded data is used in constructing prediction errors, singular value maximization is applied to maximize the singular value of the data stack, and exponential convergence of both tracking errors and estimation errors is guaranteed if regression functions of plant uncertainties are excited over a time interval such that sufficiently rich data are recorded in the data stack. Yet in this innovative design, the singular value maximization leads to an exhaustive search over all recorded data, and the requirement on the derivation of all plant states for calculating prediction errors is stringent.

This paper proposes a novel model reference composite learning control (MRCLC) strategy for a class of parametric uncertain affine nonlinear systems. In the composite learning, a modified modelling error that can utilize online recorded data is constructed as the prediction error, a second-order command filter is applied to alleviate the derivation of plant states, and both the tracking error and the prediction error are applied to update parametric estimates. It is proven that the closed-loop system achieves global exponential-like stability under interval excitation (IE) rather than PE of regression functions. Consequently, the limitations of concurrent learning are alleviated by the proposed MRCLC design. The effectiveness and superiority of this approach is verified by the application to an inverted pendulum control problem with the comparison with some existing adaptive control approaches.

The notations of this paper are relatively standard, where $\mathbb{R}$, $\mathbb{R}^+$, $\mathbb{R}^n$ and $\mathbb{R}^{n\times m}$ denote the spaces of real numbers, positive real numbers, real $n$-dimensional vectors, and $n\times m$-dimensional matrixes, respectively, $|\cdot|$ and $\|\cdot\|$ denote the absolute value and Euclidean-norm, respectively, $L_\infty$ denotes the space of bounded signals, $\Omega_c : = \{\mathbf x| \|\mathbf x\| \leq c\}$ denotes the ball of radius $c$, $\min\{\cdot\}$ and $\max\{\cdot\}$ are the minimum and maximum functions, respectively, sgn$(\cdot)$ is the sign function, rank$(A)$ is the rank of $A$, diag$(\cdot)$ is a diagonal matrix, and ${\mathcal{C}}^k$ represents the space of functions whose $k$-order derivatives all exist and are continuous, where $c \in \mathbb R^+$, $\mathbf x \in \mathbb R^n$, $A \in \mathbb{R}^{n\times m}$, and $n$, $m$ and $k$ are positive integers.

\section{Problem Formulation}

This section discusses the formulation of learning from the classical MRAC. For clear illustration, consider a class of \emph{n}th order affine nonlinear systems as follows \cite{Chowdhary2010a}:
\begin{equation}\label{eq01}
\dot{\mathbf x} = \Lambda \mathbf x + \bm b\big(f(\mathbf x) + u\big)
\end{equation}
where $\Lambda \in \mathbb R^{n\times n}$, $\bm b := [0, \cdots,0,1]^T$, $\mathbf x(t) := [x_1(t), x_2(t)$, $\cdots, x_n(t)]^T \in \mathbb R^n$ is the vector of plant states, $u(t) \in \mathbb R$ is the control input, and $f(\mathbf x): \mathbb R^n \mapsto \mathbb R$ is the $\mathcal C^1$ model uncertainty. A reference model that characterizes the desired response of the system (\ref{eq01}) is given by
\begin{equation}\label{eq02}
\dot{\mathbf x}_r = A_r \mathbf x_r + \bm b_r r
\end{equation}
with $\bm b_r := [0,\cdots, 0, b_r]^T \in \mathbb R^n$, in which $A_r \in \mathbb R^{n\times n}$ is a strictly Hurwitz matrix, $\mathbf x_r(t) := [x_{r1}(t), \cdots, x_{rn}(t)]^T \in \mathbb R^n$ is the vector of reference model states, and $r(t) \in$ $\mathbb R$ is a bounded reference signal. This study is based on the facts that $\mathbf x$ is measurable, $(\Lambda, \bm b)$ is controllable, and $f(\mathbf x)$ is linearly parameterizable such that \cite{Chowdhary2010a, Chowdhary2011, Chowdhary2014}
\begin{equation}\label{eq03}
f(\mathbf x) = W^{*T}\Phi(\mathbf x)
\end{equation}
in which $W^* \in \Omega_{c_w} \subset \mathbb R^N$ is a unknown constant parameter vector, $\Phi(\mathbf x): \mathbb R^n \mapsto \mathbb R^N$ is a known regression function vector, and $c_w \in \mathbb R^+$ is a known constant. The following definitions are given for facilitating control synthesis.

\textbf{Definition 1} \cite{Chowdhary2010a}: A bounded signal $\Phi(t) \in \mathbb R^n$ is of IE over $[t_e-\tau_d, t_e]$  if there exist constants $t_e, \tau_d, \sigma \in \mathbb R^+$ with $t_e > \tau_d$ such that $\int_{t_e-\tau_d}^{t_e}\Phi(\tau)\Phi^T(\tau)d\tau \geq \sigma I$ holds.

\textbf{Definition 2} \cite{Chowdhary2010a}: A bounded signal $\Phi(t) \in \mathbb R^n$ satisfies the PE condition if there exist constants $\sigma, \tau_d \in \mathbb R^+$ such that $\int_{t-\tau_d}^{t}\Phi(\tau)\Phi^T(\tau)d\tau \geq \sigma I$ holds, $\forall t \geq 0$.

Let $\mathbf x_{re} := [\mathbf x_{r}^T, r]^T$ be an augmented reference signal, and $\hat W \in \mathbb R^N$ be an estimate of $W^*$. Define a tracking error $\mathbf e(t) := \mathbf x(t) - \mathbf x_r(t)$, and an estimation error $\tilde{W}(t) := W^* - \hat W(t)$. Our objective is to design a proper parametric update law of MRAC such that both $\mathbf e$ and $\tilde{W}$ exponentially converge to \textbf{0} under the IE rather than the PE conation.


\section{Composite Learning Control Strategy}

\subsection{Review of Previous Results}

From \cite{Ioannou1996}, the MRAC law can be designed as follows:
\begin{align}\label{eq04}
u = &\underbrace{-\bm k_{e}^T \mathbf e} \underbrace{+ \bm k_{r}^T\mathbf x_{re}} \underbrace{-\hat W^T\Phi(\mathbf x)}\\
&\;\;\; u_{pd}\;\;\;\;\;\; u_{re}\;\;\;\;\;\;\;\;\;\; u_{ad} \notag
\end{align}
where $u_{pd}$ denotes a proportional-derivative (PD) feedback part, $u_{re}$ denotes a reference signal feedforward part, $u_{ad}$ denotes an adaptive part, $\bm k_{e} \in \mathbb R^{n}$ and $\bm k_{r} \in \mathbb R^{n+1}$ are control gains, and the design of $\bm k_{r}$ satisfies
\begin{equation}\label{eq05}
\bm b \bm k_{r}^T\mathbf x_{re} = (A_r - \Lambda)\mathbf x_{r} + \bm b_r r.
\end{equation}
Thus, one obtains the tracking error dynamics
\begin{equation}\label{eq06}
\dot{\mathbf e} = A \mathbf e + \bm b \tilde{W}^T\Phi(\mathbf x)
\end{equation}
where $A := \Lambda - \bm b\bm k_{e}^T$ is designed to be strictly Hurwitz. Thus, for any matrix $Q = Q^T > 0$, a unique solution $P = P^T > 0$ exists for the following Lyapunov equation:
\begin{equation}\label{eq07}
A^TP + PA = -Q.
\end{equation}

Let the adaptive law of $\hat W$ be as follows:
\begin{equation}\label{eq08}
\dot{\hat W} = \mathcal P(\hat W, \gamma\mathbf e^TP\bm b \Phi(\mathbf x))
\end{equation}
where $\gamma \in \mathbb R^+$ is a learning rate, and $\mathcal P(\hat W, \bullet)$ is a projection operator given by \cite{Ioannou1996}
\begin{equation*}
\mathcal P(\hat W, \bullet) = \left\{\begin{array}{l}
\bullet - {\hat W \hat W ^T \cdot \bullet}/{\|\hat W \|^2}, \\
\;\; \mathrm{if} \; \|\hat W\| \geq c_w\; \& \; \hat{W}^T \cdot \bullet > 0 \\
\bullet, \; \mathrm{otherwise}
\end{array}.\right.
\end{equation*}
Choose a Lyapunov function candidate
\begin{equation}\label{eq09}
V(\mathbf z) = \mathbf e^TP\mathbf e/2 + \tilde{W}^{T}\tilde{W}/(2\gamma)
\end{equation}
with $\mathbf z := [\mathbf e^T, \tilde{W}^{T}]^T \in \mathbb R^{n+N}$ for the closed-loop dynamics composed of (\ref{eq06}) and (\ref{eq08}). It follows from the classical MRAC result of \cite{Ioannou1996} that if $\hat{W}(0) \in \Omega_{c_w}$ and $\Phi(\mathbf x)$ is of PE, then the closed-loop system (\ref{eq06}) with (\ref{eq08}) achieves \emph{global exponential stability} in the sense that both the tracking error $\mathbf e(t)$ and the estimation error $\tilde{W}(t)$ converge to \textbf{0}.

To remove the requirement of the PE condition on $\Phi(\mathbf x)$ for parameter convergence in \cite{Ioannou1996}, a concurrent learning law of $\hat W$ is proposed in \cite{Chowdhary2010a} as follows:
\begin{equation}\label{eq10}
\dot{\hat W} = \; \mathcal P\Big(\hat W, \gamma\mathbf e^TP\bm b \Phi(\mathbf x) + \gamma\sum_{j=1}^{p} \epsilon_j \Phi^T(\mathbf x_j)\Big)
\end{equation}
where $j$ is a certain epoch, $p \geq N$ is the number of stored data, and $\epsilon_j := \tilde{W}_j^T\Phi(\mathbf x_j)$ denotes a modelling error which is regarded as the prediction error calculated by
\begin{equation}\label{eq11}
\epsilon_j = \bm b^T(\dot{\mathbf x}_j - \Lambda \mathbf x_j - \bm b u_j) - \hat W_j^T\Phi(\mathbf x_j)
\end{equation}
where $\mathbf x_j$, $u_j$ and $\hat W_j$ are online recorded data of $\mathbf x$, $u$ and $\hat W$ at the epoch $j$, respectively. Let $Z :=[\Phi(\mathbf x_1), \Phi(\mathbf x_2), \cdots, \Phi(\mathbf x_p)]$ $\in \mathbb R^{p \times N}$ be a dynamic data stack. It is shown in the concurrent learning MRAC approach of \cite{Chowdhary2010a} that if $\hat{W}(0)\in$ $\Omega_{c_w}$ and rank$(Z) = N$ on $t \geq T_e$, then the closed-loop system (\ref{eq06}) with (\ref{eq10}) achieves \emph{global exponential-like stability} in the sense that both the tracking error $\mathbf e$ and the estimation error $\tilde{W}$ converge to \textbf{0} on $t \geq T_e$. Yet in the approach of \cite{Chowdhary2010a}, fixed-point smoothing must be applied to estimate $\dot{\mathbf x}_j$ in (\ref{eq11}) such that the prediction error $\epsilon_j$ is calculable, and singular value maximization should be applied to exhaustively search the data stack $Z$ to maximize its singular value.

\subsection{Composite Learning Control Design}

Define a modified modelling error $\bm\epsilon(t) := \Theta_e\tilde{W}(t)$ as the prediction error, in which
\begin{equation}\label{eq12}
\Theta_e := \left\{\begin{array}{l}
\Theta(t) \;\;\;\; \mathrm{for}\;t < T_e \;\mathrm{if}\; \Theta(t) < \sigma I \\
\Theta(T_e) \;\mathrm{for}\;t \geq T_e \;\;\mathrm{if}\; \Theta(T_e) \geq \sigma I\\
\end{array}\right.
\end{equation}
with $\sigma = \max_{t\geq 0}\{\sigma_r(t)\}$ and
\begin{equation}\label{eq13}
\Theta(t) := \int_{t-\tau_d}^{t} \Phi(\mathbf x(\tau)) \Phi^T(\mathbf x(\tau)) d\tau
\end{equation}
in which $\tau_d \in \mathbb R^+$ is an integral duration, $\sigma_r(t) \in \mathbb R^+$ is the minimal singular value of $\Theta(t)$, and $T_e \geq \tau_d$ is the time when $\sigma_r(t)$ reaches its maximum\footnotemark[1]. Then, a composite learning law of $\hat{W}$ is designed as follows:
\begin{equation}\label{eq14}
\dot{\hat{W}} = \gamma \;\mathrm{proj}\big(\mathbf e^TP\bm b \Phi(\mathbf x) + k_w \bm\epsilon\big)
\end{equation}
in which $k_w \in \mathbb R^+$ is a weight factor.

\footnotetext[1]{The presence of $T_e$ and $\sigma$ here is only used in the subsequent performance analysis and their values can only be gotten at the control process.}

From $\bm\epsilon = \Theta_e\tilde{W}$ and (\ref{eq12}), the calculation of $\bm\epsilon$ can follow the calculation of $\Theta \tilde{W}$ as follows:
\begin{equation}\label{eq15}
\Theta(t)\tilde{W}(t) = \Theta(t) W^* - \Theta(t)\hat{W}(t)
\end{equation}
where $\Theta\hat{W}$ is directly obtainable by (\ref{eq13}) and (\ref{eq14}), and $\Theta W^*$ is not directly obtainable. Multiplying both sides of (\ref{eq06}) by $\Phi(\mathbf x(t))$, int\-egrating the resulting equality over $[t-\tau_d, t]$ and making some transformations, one gets
\begin{equation}\label{eq16}
\Theta(t) W^* = \int_{t-\tau_d}^{t} \Phi(\mathbf x)(\dot{e}_n + \hat W^T\Phi(\mathbf x)-\bm b A \mathbf e)d\tau
\end{equation}
in which the time variable $\tau$ is omitted in the above integral part. Since $\dot{e}_n$ is unavailable, a second-order linear filter with unit gain is implemented as follows \cite{Hu2013}:
\begin{equation}\label{eq17}
\left\{\begin{array}{l}
\dot{\hat e}_n = {\hat e}_{n+1}\\
\dot{\hat e}_{n+1} = -2\zeta\omega {\hat e}_{n+1} + \omega^2(e_n - {\hat e}_n)
\end{array}\right.
\end{equation}
with ${\hat e}_n(0) = e_n(0)$ and ${\hat e}_{n+1}(0) = 0$, where $\omega \in \mathbb R^+$ is the natural frequency, $\zeta \in \mathbb R^+$ is the damping factor, and ${\hat e}_n$ and ${\hat e}_{n+1}$ are estimates of ${e}_n$ and ${\dot e}_{n}$, respectively. The integral in (\ref{eq16}) effectively reduces the influence of measurement noise on the calculation of $\Theta W^*$. Thus, $\omega$ in (\ref{eq17}) can be made sufficiently small so that ${\hat e}_{n+1} \approx {\dot e}_{n}$. For simplifying analysis, assume that ${\hat e}_{n+1} = {\dot e}_{n}$ as in \cite{Hu2013}. The reasonability of this assumption will also be verified in the subsequent simulations. The following theorem shows the main result of this study.

\textbf{Theorem 1}: Consider the system (\ref{eq01}) driven by the control law (\ref{eq04}) with (\ref{eq14}), where the control gain $\bm k_{r}$ is designed to satisfy (\ref{eq05}), and the control gain $\bm k_{e}$ is selected to make $A$ in (\ref{eq06}) strictly Hurwitz. If $\hat{W}(0) \in \Omega_{c_w}$ and $\Theta(T_e) \geq \sigma I$ are satisfied with $c_w, \sigma \in \mathbb R^+$ and $T_e$ $\geq$ $\tau_d$, then the closed-loop system composed of (\ref{eq06}) and (\ref{eq14}) achieves \emph{global exponential-like stability} in the sense that all closed-loop signals are bounded for all $t \geq 0$ and both the tracking error $\mathbf e(t)$ and the estimation error $\tilde{W}(t)$ exponentially converge to \textbf{0} on $t \geq T_e$.

\emph{proof:} Firstly, consider the control problem at $t \in [0, \infty)$. Choose the Lyapunov function candidate $V$ in (\ref{eq09}) for the closed-loop system constituted by (\ref{eq06}) and (\ref{eq14}). The time de\-rivative of ${V}$ along (\ref{eq06}) is as follows:
\begin{equation}\label{eq21}
\dot V = -\mathbf e^TQ\mathbf e/2 + \tilde{W}^T \big(\mathbf e^T P \bm b \Phi(\mathbf x) - \dot{\hat{W}}\big)
\end{equation}
where (\ref{eq07}) is utilized to obtain (\ref{eq21}). Applying (\ref{eq14}) to (\ref{eq21}), noting $\hat{W}(0) \in \Omega_{c_w}$ and $\bm\epsilon = \Theta_e\tilde{W}$, and using the pro\-jection operator results in \cite{Ioannou1996}, one obtains
\begin{equation}\label{eq22}
\dot V \leq -\mathbf e^TQ\mathbf e/2 - k_w \tilde{W}^T\Theta_e\tilde{W},\forall t \geq 0.
\end{equation}
Noting $Q > 0$ and $\Theta_e \geq 0$, one gets $\dot V(t) \leq 0$, $\forall t\geq 0$, which implies that the closed-loop system is stable in the sense that $\mathbf e, \tilde{W} \in L_\infty$. Since  $\dot V(t) \leq 0$ is satisfied $\forall \mathbf x(0) \in \mathbb R^n$ and $V$ in (\ref{eq09}) is radially unbounded (i.e. $V(\mathbf z) \rightarrow \infty$ as $\mathbf z \rightarrow \infty$), the stability is global. Using $\mathbf e, \tilde{W} \in L_\infty$, one also obtains $\mathbf x, \hat{W}, \Phi, \bm\epsilon, u \in L_\infty$ from their definitions. Thus, all closed-loop signals are bounded for all $t\geq 0$.

Secondly, consider the control problem at $t \in [T_e, \infty)$. since there exist $\sigma \in \mathbb R^+$ and $T_e \geq \tau_d$ such that $\Theta(T_e) \geq \sigma I$, i.e. the bounded signal $\Phi(\mathbf x(t))$ is exciting over $t \in [T_e-\tau_d, T_e]$, it is obtained from (\ref{eq22}) that
\begin{equation}\label{eq23}
\dot V \leq -\mathbf e^TQ\mathbf e/2 - k_w\sigma \tilde{W}^T\tilde{W}, \forall t \geq T_e.
\end{equation}
It follows from (\ref{eq09}) and (\ref{eq23}) that
\begin{align}\label{eq24}
\dot V(t) & \leq - k_s V(t), \forall t \geq T_e
\end{align}
where $k_s := \min\{\lambda_{\min}(Q)/\lambda_{\max}(P), 2\gamma k_w\sigma\} \in \mathbb R^+$, which implies that the closed-loop system has global exponential-like stability in the sense that both $\mathbf e$ and $\tilde{W}$ exponentially converge to \textbf{0} as long as $t \geq T_e$. \hfill $\Box$

\section{Application to Inverted Pendulum} \label{Example}

\begin{figure*}[!t]
\centering
\subfigure[]{\includegraphics[width = 3.5in]{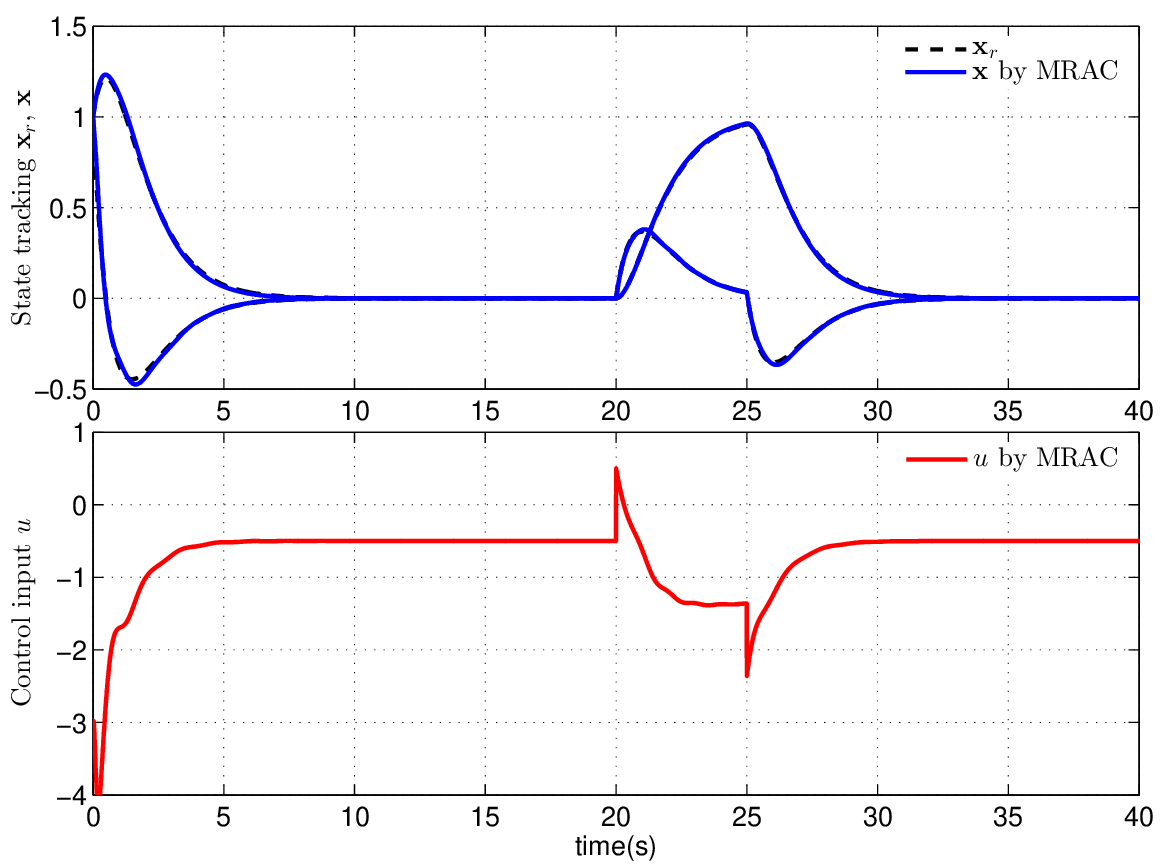}}\hfill
\subfigure[]{\includegraphics[width = 3.5in]{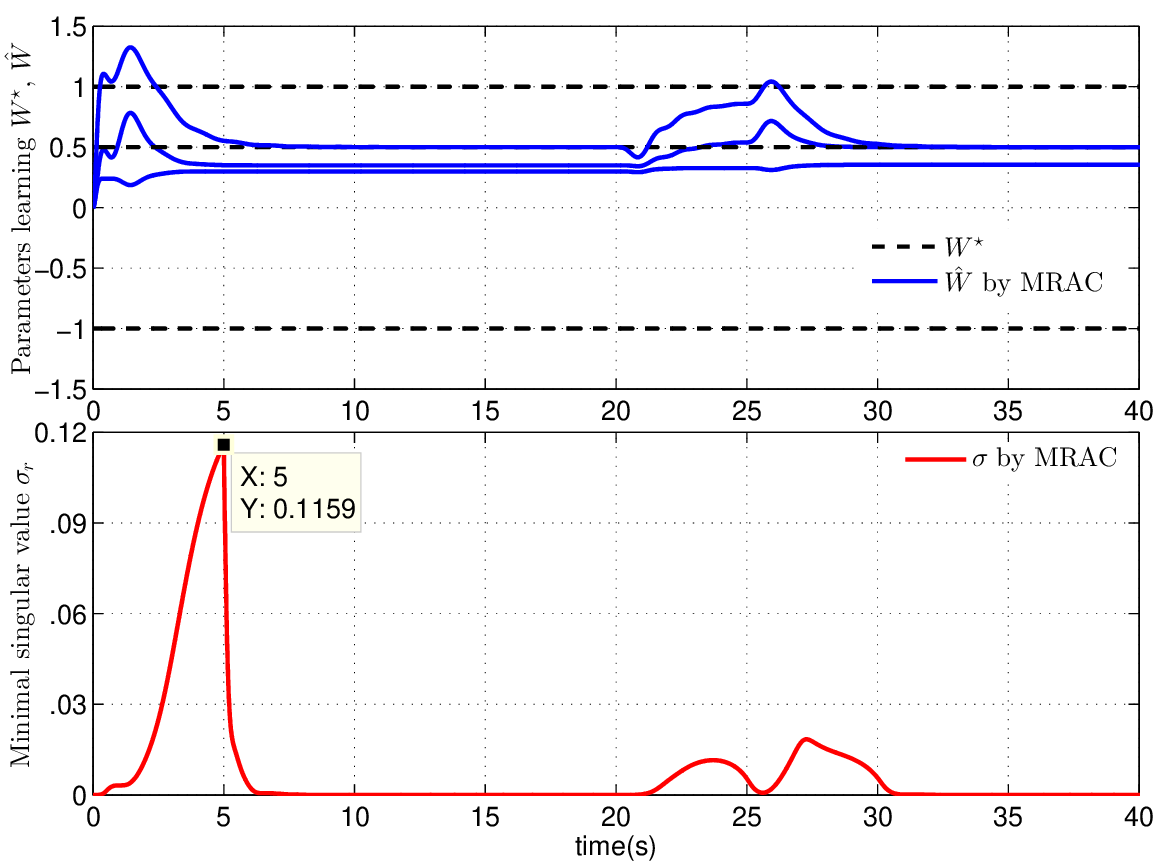}}\hfill
\subfigure[]{\includegraphics[width = 3.5in]{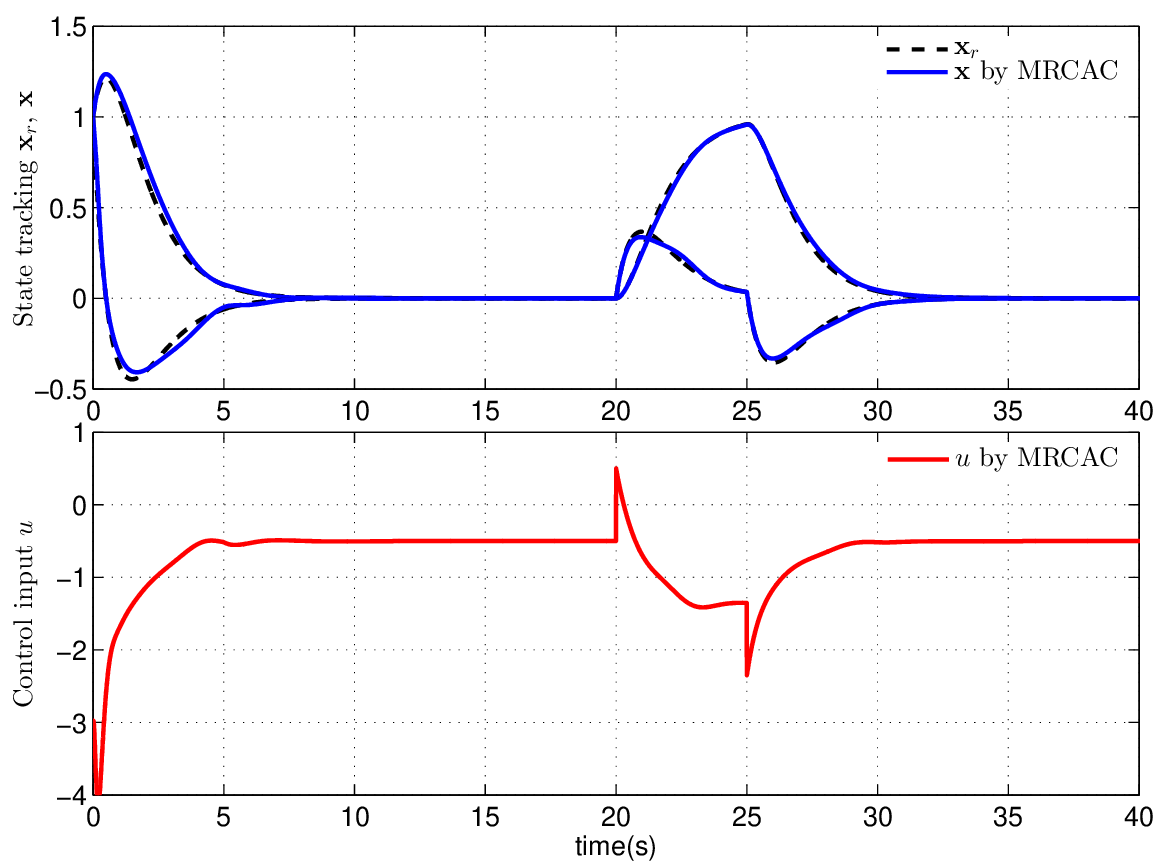}}\hfill
\subfigure[]{\includegraphics[width = 3.5in]{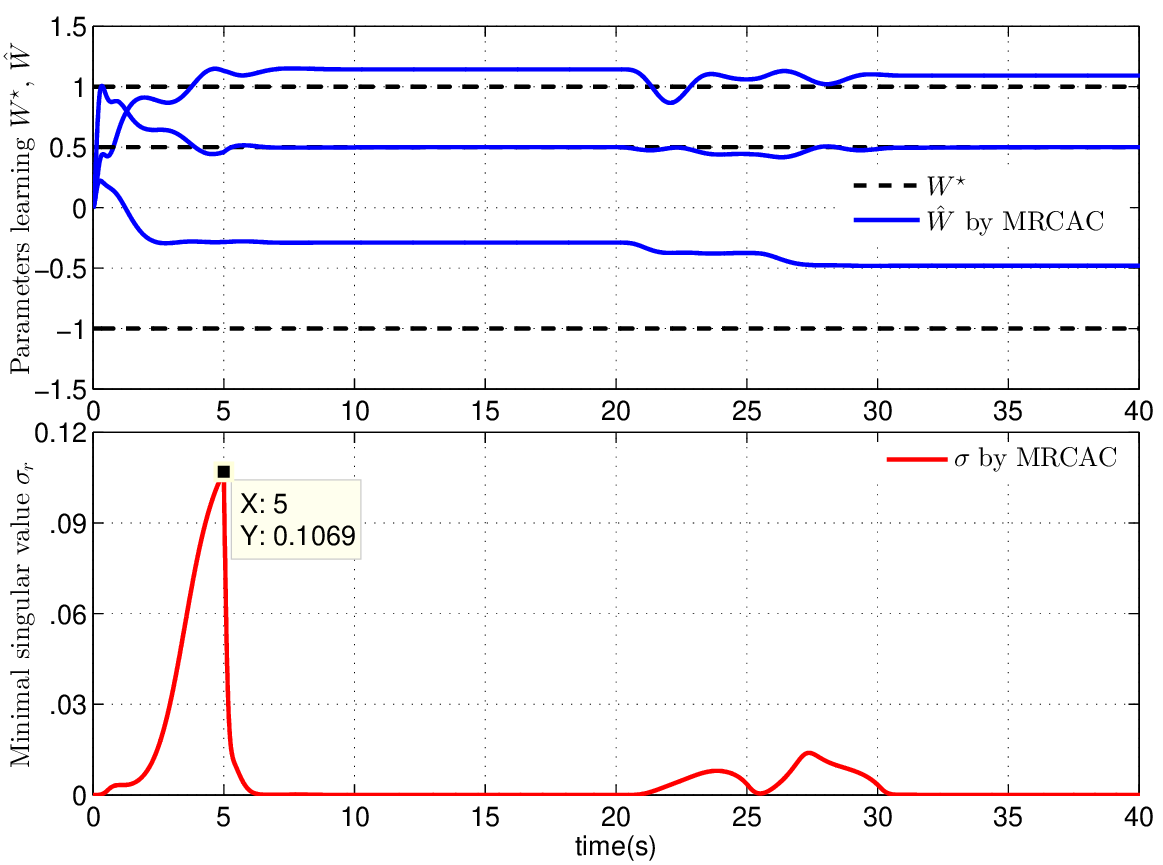}}\hfill
\subfigure[]{\includegraphics[width = 3.5in]{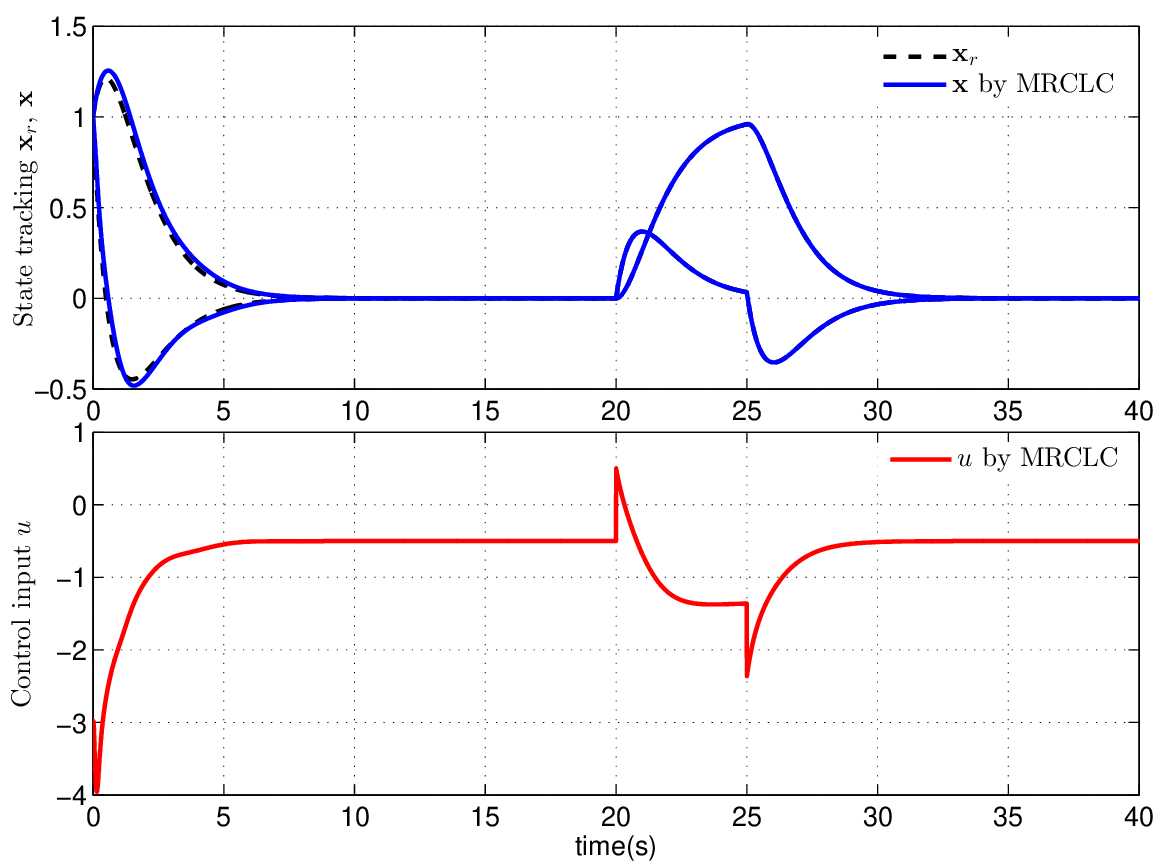}}\hfill
\subfigure[]{\includegraphics[width = 3.5in]{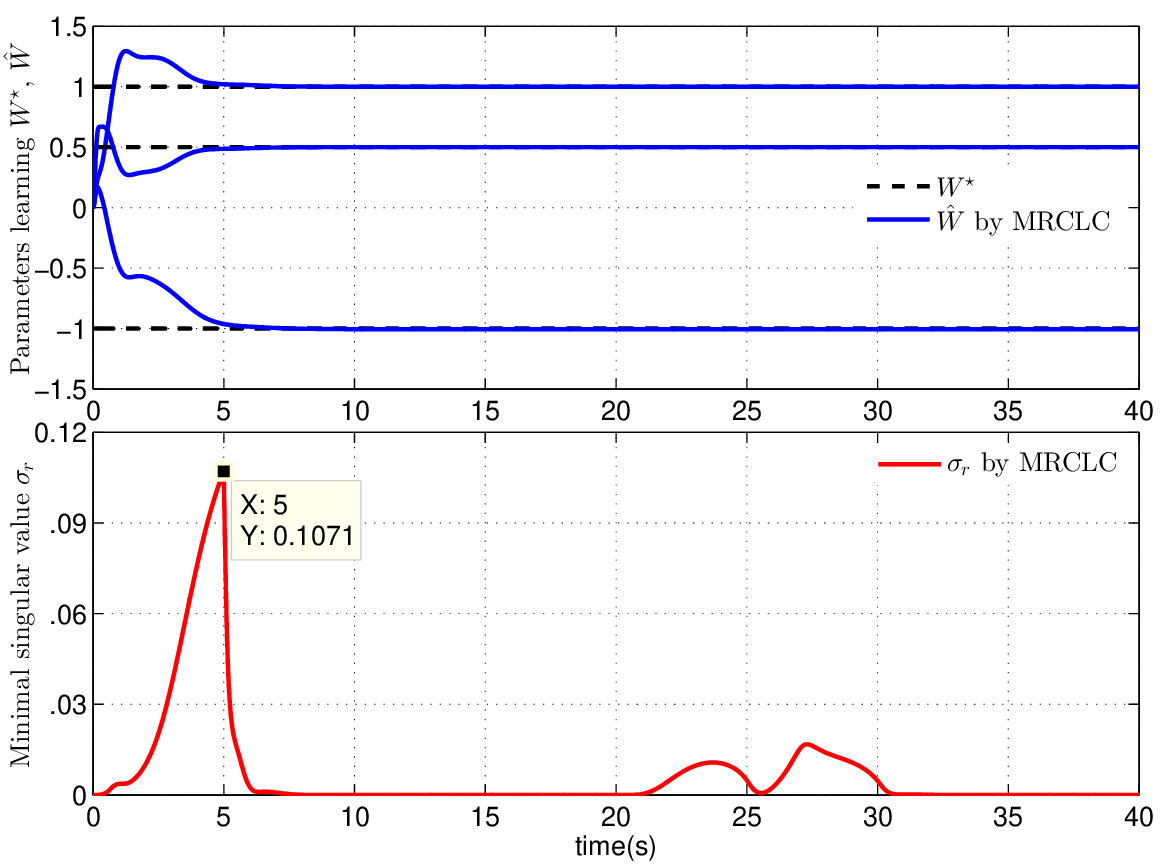}}\\
\caption{Simulation trajectories by all controllers. (a) Control performance by the MRAC. (b) Learning performance by the MRAC. (c) Control performance by the MRCAC. (d) Learning performance by the MRCAC. (e) Control performance by the MRCLC. (f) Learning performance by the MRCLC.}
\label{fig01}
\end{figure*}

Consider the following inverted pendulum model \cite{Chowdhary2010a}:
\begin{equation*}
\dot{\mathbf x} = \left[\begin{array}{cc}
                             0 & 1 \\
                             0 & 0 \\
                           \end{array}\right]
 \mathbf x + \left[\begin{array}{c}
                    0 \\
                    1
                  \end{array}\right]
 (W^{*T}\Phi(\mathbf x) + u)
\end{equation*}
in which $W^* = [1, -1, 0.5]^T$ and $\Phi(\mathbf x) = [\sin x_1, |x_2|x_2$, $e^{x_1x_2}]^T$. The reference model is given as follows:
\begin{equation*}
\dot{\mathbf x}_r = \left[\begin{array}{cc}
                             0 & 1 \\
                             -1 & -2 \\
                           \end{array}\right] \mathbf x_r +
                           \left[\begin{array}{c}
                    0 \\
                    1
                  \end{array}\right] r.
\end{equation*}
For simulations, set $\mathbf x(0) = \mathbf x_r(0) = [1, 1]^T$, $r(t) = 1$ while $t \in [20, 25)$, and $r(t) = 0$ while $t \in [0, 20) \cup [25, \infty)$ \cite{Chowdhary2010a}.

The parameters selection of the proposed control law (\ref{eq04}) with (\ref{eq14}) is based on that of \cite{Chowdhary2010a}, where the details are given as follows: firstly, solve (\ref{eq05}) to obtain $\bm k_{r} = [-1, -2, 1]^T$; sec\-ondly, select $\bm k_{e} = [1.5, 1.3]^T$ so that $A$ is strictly Hurwitz; thirdly, solve (\ref{eq07}) with $Q = 10I$ to obtain $P$; fourthly, set $\tau_d$ $=$ 5 s in (\ref{eq13}); fifthly, set $\gamma =$ 3.5, $k_w =$ 6 and $c_w =$ 5 in (\ref{eq14}); finally, set $\omega =$ 100 and $\zeta =$ 0.7 in (\ref{eq17}).

Simulations are carried out in MATLAB software running on Windows 7, where the solver is set to be fixed-step ode 5 with the step size being 0.001 s and other settings being default values. The conventional MRAC in \cite{Ioannou1996} and the model reference CAC (MRCAC) with \emph{Q}-modification in \cite{Volyanskyy2010} are selected as baseline controllers, where the reference models and shared parameters of all controllers are set to be the same for fair comparisons. Simulation trajectories by all controllers are depicted Fig. \ref{fig01}. For the control performance, it is shown that the plant state $\mathbf x$ follows its desired signal $\mathbf x_r$ closely with a smooth control input $u$ for all controllers, the MRCAC achieves the worst tracking accuracy [see Fig. \ref{fig01}(c)], and the proposed MRCLC achieves the best tracking accuracy [see Fig. \ref{fig01}(e)]. For the learning performance, it is observed that IE occurs with $\sigma$ rising at $t = 5$ s in this case, the MRAC does not show any parameter convergence [see Fig. \ref{fig01}(b)], the MRCLC shows better parameter estimation than the MRAC but still does not achieve parameter convergence [see Fig. \ref{fig01}(d)], and the proposed MRCLC achieves fast and accurate parameter estimation even the IE is short and weak [see Fig. \ref{fig01}(f)].

\section{Conclusion}

In this paper, a MRCLC strategy has been successfully developed for a class of parametric uncertain affine nonlinear systems such that parameter convergence can be guaranteed by the IE rather than PE condition. The proposed approach has also been applied to an inverted pendulum model, where superior control and learning performances have been demonstrated compared with the conventional MRAC and the MRCAC with Q-modification. 
Further work would focus on the extension of the composite learning to wider classes of nonlinear systems such as multi-input multi-output affine nonlinear systems \cite{Pan2013a} and strict-feedback nonlinear systems \cite{Pan2015}.


\section*{Acknowledgment}

This work was supported in part by the National Research Foundation of Singapore under Grant NRF2014NRF-POC001-027, in part by the Biomedical Engineering Programme, Age\-ncy for Science, Technology and Research (A*STAR), Sing\-apore under Grant 1421480015, and in part by AFR and FNR programs, Luxembourg.



%

\end{document}